\documentclass{article}
\usepackage{graphicx} 
\usepackage{lineno}
\usepackage{subcaption}
\usepackage{amsmath}
\usepackage{algorithm}
\usepackage{booktabs}
\usepackage{algpseudocode}
\usepackage[acronym]{glossaries}
\usepackage{setspace}
\usepackage{authblk}
\usepackage{comment}
\usepackage{manyfoot}
\usepackage[short]{optidef}
\usepackage{lineno}
\usepackage{textcomp}
\usepackage{multirow}
\usepackage{caption}
\usepackage{amssymb}
\usepackage{nomencl}
\usepackage[most]{tcolorbox}
\usepackage{adjustbox}
\usepackage{booktabs}
\usepackage{etoolbox}
\usepackage{algpseudocode}
\usepackage{algorithm}
\usepackage{tikz}
\usetikzlibrary{shapes.geometric, arrows.meta}
\newcommand{\mathz}{\ooalign{$z$\cr\hfil\rule[.5ex]{.2em}{.06ex}\hfil\cr}}

\begin{document}

\title{Efficient Training for Optical Computing}

\author[1]{Manon P. Bart}
\author[1]{Nick Sparks}
\author[1]{Ryan T. Glasser}
\affil[1]{Department of Physics and Engineering Physics, Tulane University}

\date{}

\maketitle

\begin{abstract}

Diffractive optical information processors have demonstrated significant promise in delivering high-speed, parallel, and energy efficient inference for scaling machine learning tasks. Training, however, remains a major computational bottleneck, compounded by large datasets and many simulations required for state-of-the-art classification models. The underlying linear transformations in such systems are inherently constrained to compositions of circulant and diagonal matrix factors, representing free-space propagation and phase and/or amplitude modulation of light, respectively. While theoretically established that an arbitrary linear transformation can be generated by such factors, only upper bounds on the number of factors exist, which are experimentally unfeasible. Additionally, physical parameters such as inter-layer distance, number of layers, and phase-only modulation further restrict the solution space. Without tractable analytical decompositions, prior works have implemented various constrained minimization techniques. As trainable elements occupy a small subset of the overall transformation, existing techniques incur unnecessary computational overhead, limiting scalability. In this work, we demonstrate significant reduction in training time by exploiting the structured and sparse nature of diffractive systems in training and inference. We introduce a novel backpropagation algorithm that incorporates plane wave decomposition via the Fourier transform, computing gradients across all trainable elements in a given layer simultaneously, using only change-of-basis and element wise multiplication. Given the lack of a closed-form mathematical decomposition for realizable optical architectures, this approach is not only valuable for machine learning tasks but broadly applicable for the generation of arbitrary linear transformations, wavefront shaping, and other signal processing tasks.

\end{abstract}

Decades of research has prefaced the ubiquity of machine learning in our present lives. Deep learning in particular has proven to be a multi-disciplinary tool, with research developments and industry applications in speech \cite{SpeechRec} and object recognition \cite{VisRec}, biomedical applications \cite{Health0,Health1}, marketing and advertisement \cite{Market}, and stock market analysis \cite{Stock}, among others. Despite the advancements in parallelism and processing power of hardware in recent years, modern technologies are approaching scaling limitations and drastic energy consumption due to the massive computing power and computational complexity needed for statistical inference \cite{EComp0,EComp1}. These factors have serious environmental implications, especially during training and hyperparameter tuning \cite{EnergyCon}. Consequently, the field is increasingly shifting toward more sustainable, low-power approaches to machine learning. While early proposals of diffractive optical processors and machine learning \cite{Farhat1985,Porcello1960,Lugt1964} were largely overshadowed by the rise of deep neural networks (DNNs) coupled with rapid advances in graphical processing units (GPUs) \cite{OpticsHistory}, renewed interest in optical processing can be attributed to the physical architecture's low power consumption, high speed and bandwidth, and intrinsic parallelism \cite{OpticsBackground, AnaOptComp, ONN_R0}.   

While their are many ways to process optical information, such as hybrid opto-electronics, silicon photonic circuits, and diffractive surfaces or cascaded phase masks \cite{ProcO0,ProcO1,ProcO2,ProcO3}, all of these processors leverage certain light matter interactions to perform linear transformations which are computationally expensive on current computing architectures \cite{Goodman,OLT0,OLT3}. Optical information processing has shown functionality in the case of free space communication \cite{FSO}, optical imaging \cite{OApp_Im0}, optical interconnections \cite{OApp_IC}, biomedical applications \cite{OApp_Bio0,OApp_Bio1,OApp_Bio2}, solving complex equations \cite{OApp_Eq0,OApp_Eq1}, and demonstrations of quantum linear operations \cite{OApp_Q0,OApp_Q1}, among others. Optical machine learning has matured in recent years with many analog optical neural network architectures such as single photon optical vector matrix multiplication, massively parallel diffractive neural networks, optical reservoir computing, and low to zero-power convolutional neural networks \cite{OLTNN0, ProcONN1, CONN0, CONN5, ONN_R0, Multiplex0}. Diffraction-based optical architectures are particularly appealing for processing multi-dimensional data. This is due to the multiple degrees of freedom such as phase, amplitude \cite{ONN2,ONN3}, polarization \cite{ONN_p}, which, unlike integrated photonics, are retained at each layer, leading to the ability to recover information to mitigate errors in training and implementation \cite{DONN_R0, BPONN1}. Further, task specific processes which rely on real time correction suffer from low latency during pre-processing, whereas direct encoding can be done on free-space optical processors \cite{OApp_Im0,DirectEncode0,DirectEncode1}.

For diffraction-based optical processors, various physical architectures have been explored to implement linear transformations such as convolutions \cite{CONN5,Goodman1966} or optical vector-matrix multiplication (OVMM) \cite{ProcO2,Goodman1978}. In recent years, cascaded phase masks or diffractive elements have gained significant traction as a scalable approach to machine learning tasks \cite{OLT0,OLT3}. Information is typically encoded in the phase and amplitude channels of light and trainable phase masks, defined by a diagonal matrix, modulate the complex field of light at each layer. When cascaded with free-space propagation, represented as circulant matrices, these systems can realize arbitrary linear transformations \cite{OLTBase,OLTCirc1} with extremely low power consumption \cite{OLT0,OLT3,OLT2} and non-destructive, unitary transformations that preserve optical information \cite{OLT1}. Unlike OVMM, where direct encoding of the matrix representing the linear transformation is encoded into a diffractive element \cite{ProcO2, OLTNN0, OLTNN1}, decomposing a linear transformation into circulant and diagonal factors efficiently remains non-trivial \cite{OLTCirc0,OLTCirc1}. Despite proof that a composition of circulant and diagonal matrices can generate an arbitrary linear transformation \cite{OLTBase}, the generation of these matrices is constructive, and involves solving a structured system of polynomial equations \cite{OLTCirc0}. An upper bound of $2\text{N}-1$ alternating circulant and diagonal blocks for an exact decomposition of an $\text{N} \times \text{N}$ linear transformation has been established, but this is experimentally and computationally prohibitive \cite{OLTCirc0}. In addition, constraints such as task-specific parameters for physical setups (i.e. wavelength, sampling, inter-layer distances) as well as phase-only modulation and finite resolution make finding a suitable decomposition for a small number of circulant and diagonal factors a considerable challenge. 

For machine learning tasks, determination of an exact, unique decomposition is unnecessary, only requiring a sufficiently expressive optical architecture. Research has focused on strategies to address optical nonlinearities \cite{ONN_nl2,ONN_nl1,ONN_nl3,ONNNL2} and architectural complexity \cite{ONN0,ONN2,ONN3} in order to create models that approach the performance of state-of-the-art electronic machine learning models. A major bottleneck in scaling these efforts, however, remains training them. To address this, several strategies have been proposed using constrained minimization. A matrix pseudoinverse-based synthesis was utilized for single-layer complex linear transformations, but was not clearly generalizable to multi-layer architectures required for expressive linear transformations \cite{OLT0}. A finite difference method has been proposed; however, it requires performing forward propagation twice for each individual trainable parameter, which constrains the scalability that optical architectures offer \cite{BPOpt0}.  Data-driven approaches using backpropagation, which calculate the gradients and error in trainable parameters with respect to a cost function and updates them to reach a local minima, have emerged as the most effective strategy \cite{BP}. Nonetheless, these approaches still present several challenges. Unlike traditional ANNs, where a weight corresponds to a scalar mapping between an input and an output element, each trainable element in a DONN affects the output globally. Further, optical architectures necessitate multiple cascaded transformations to achieve arbitrary linear mappings, increasing training times \cite{OLT1}. For deep DONNs, the difficulty compounds, as architectural optimization, limited trainable elements, and thousands of images are required to successfully perform machine learning tasks. 

To address the persistent long training times of backpropagation, DONN architectures, modeled \textit{in-silico} using the Rayleigh–Sommerfeld diffraction integral, have been modified to adopt the angular spectrum method for feed forward propagation, leveraging the simplification of convolutions into element-wise multiplication in the Fourier domain. This enables the use of auto-differentiation libraries like TensorFlow (Google) to approximate gradients, often outperforming other approaches as long as certain conditions, like the number of diffractive surfaces or field-of-view (FOV) size, are satisfied \cite{IC_D0,OLT0}. 
These methods, however, remain time-intensive, with reported training durations on GPUs ranging from 6 to 48 hours for a single training implementation \cite{BPONN1,ONN0,ONN3,ONNTimes}. In addition, derivation of the gradients for the angular spectrum method have not been established in literature. More recently, \textit{in situ} methods have been proposed. These methods can perform optical forward propagation, while handling backpropagation electronically.  These reduce both training time and the simulation–experiment gap by accounting for experimental imperfections  \cite{BPONN0}. However, this process requires camera imaging and phase retrieval techniques to determine the phase and amplitude of the light at each layer. In order to scale DONNs, current models still necessitate computation of gradients \textit{in-silico}, which remains bounded by computer speed and memory \cite{DONN_R0}. 

In this work, we derive a novel method to calculate gradients based on plane wave decomposition, with a large decrease in computational time as compared to auto-differentiation. To our knowledge, this is the first Fourier-decomposition-based backpropagation algorithm that fully incorporates the physics of optical propagation. As the gradient is a high dimensional vector, unwanted computational time is not wasted on calculating the full Jacobian, but rather accounted for intrinsically through Fourier decomposition, allowing for direct adjustment of relevant physical parameters using only element-wise operations in the Fourier domain. Unlike finite-difference methods, which assess one weight at a time, our algorithm evaluates the influence of all weights on the cost function simultaneously in a given layer. This approach can also be used in conjunction with methods which rely on physics-aware backpropagation to adjust the constrained trainable parameters. Beyond DONN training, this is broadly applicable to optical computing tasks requiring arbitrary linear transformations or signal processing where circulant and diagonal matrices are used.

\section*{Results}
\subsection*{Diffractive Optical Neural Network Architecture} \label{sec:arch}
\begin{figure}[htbp!]
    \centering
    \includegraphics[width=\linewidth]{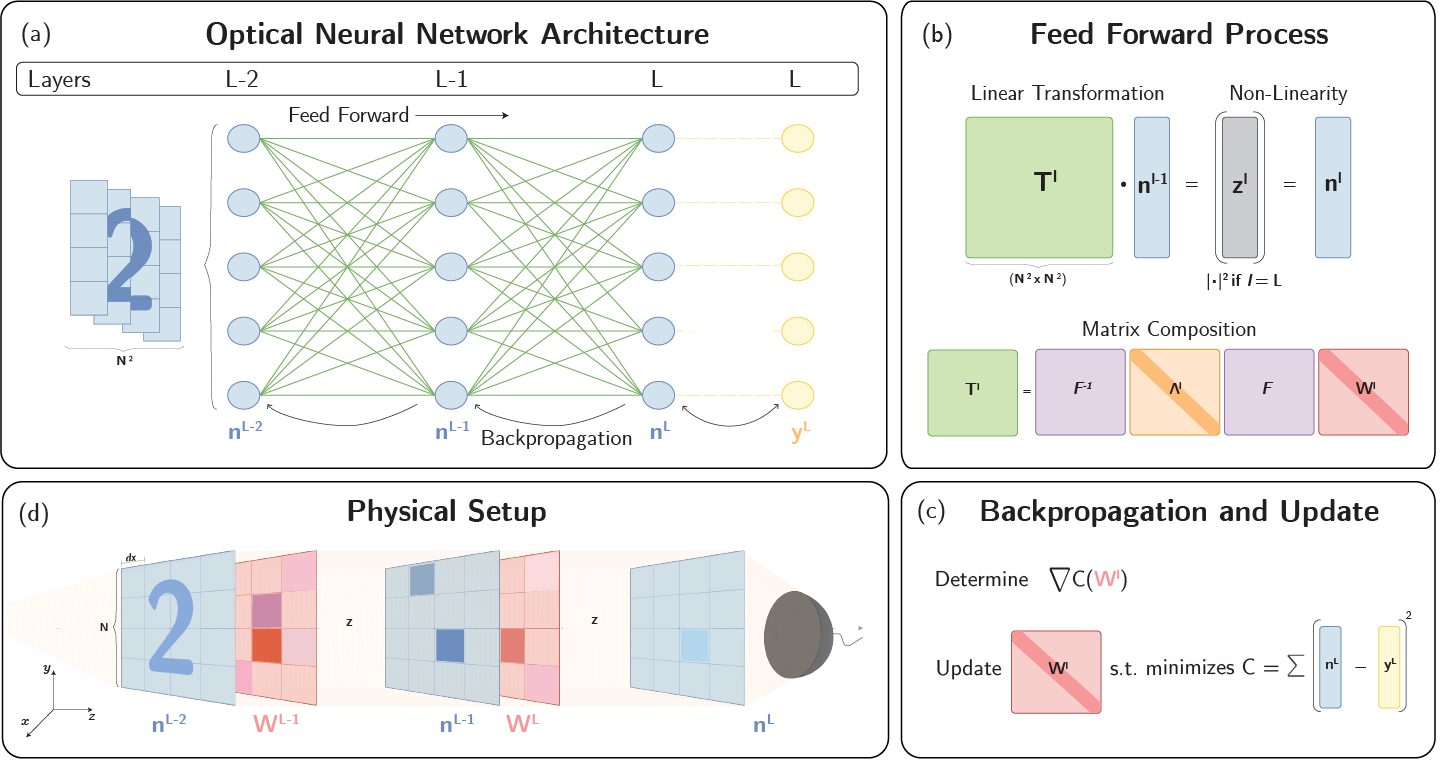}
    \caption{ Optical neural network architecture for L = 3 layers of complex neurons. Input data is encoded into the complex neurons phase and/or amplitude channels at $\text{n}^{l =0}_k$, and imaged at $n^{l = \text{L}}_i$. (a) The complex neurons, $\text{n}^{l}$ are represented in blue by their column wise vectorization of dimension ($\text{N}^2 \times1$ ). The target output, $\text{y}^l$, is shown in yellow. The linear transformation is defined by the green connections between neurons. (b) Feed forward process. The linear transformations are described by transformation matrix, T$^l$,  of dimension  (N$^2 \times$ N$^2$). This can be defined by composition of the trainable weight matrix, $\hat{W}^l = \text{diag}(e^{i\phi^l})$, followed by free space propagation. The non-linearity at layer $l = \text{L}$ is defined by the magnitude squared of the vector. (c) The backpropagation process involves determining the gradient of the cost function, $\nabla C (\hat{W}^l)$, which is defined as mean-squared-error, with respect to the trainable weights, $W^l$. (d) Analogous physical setup. Complex neurons and weights, which are sampled values in the $(x,y)$ plane of dimension ($\text{N}\times \text{N}$) are shown in blue and red, respectively. Layers are defined by distance from each previous plane $l-1$, where the final layer is given by the imaged intensity at n$^{\text{L}}$.}
    \label{fig:ONNArch}
\end{figure}

At their core, neural networks consist of linear transformations and non-linear activations to capture complex patterns and relationships within data. Layers of these transformations enable neural networks to fit, generalize, and model complex data distributions with high precision. For a typical ANN, a neuron at any layer $l$, denoted $n^l_{i}$,  undergoes a linear transformation with a trainable weight matrix and bias vector. If we express the neurons in vectorized notation, the transformation at layer $l-1$ corresponds to $
    \vec{z}\,^l = \hat{W}^l\vec{n}^{l-1} + \vec{b}^l$, followed by a non-linear transformation, $f$, $\vec{n}^l = f(\vec{z}\,^l)$, resulting in the output neurons at layer $l$ \cite{DL_nielsen}. In a DONN, the neurons, as well as their linear and non-linear transformations, adhere to the intrinsic physical properties of optical systems. For a wave traveling in the $+ \mathz $ direction, each layer of a DONN describes the complex wavefront in the $(x,y)$ plane a distance, $\mathz $, away from the source. An optical neuron, $n^l_i = |n^l_i| e^{i\phi_i^l}$ , corresponds to the amplitude and phase at a discrete point on the wavefront in spatial positions $(x,y)$. A linear transformation, $\hat{T}^l$, between two layers can be represented by amplitude and/or phase modulation followed by the propagation of the optical field from one layer to the next in the $+\mathz $ direction, which can be modeled as $\vec{z}\,^l  =  \hat{T}^l \vec{n}^{\text{l-1}}$, where $\vec{n}^{l-1}$ corresponds to the complex neurons in the $(x,y) $ plane vectorized in the input layer of dimension ($\text{N}^2 \times 1$),  $\vec{z}\,^l $ corresponds to the output following the linear transformation of dimension ($\text{M}^2 \times 1$) , and $\hat{T}^l$ of dimension ($\text{N}^2 \times \text{M}^2$). The following architecture can be visualized by Figure \ref{fig:ONNArch}.

Following a linear transformation, a non-linearity, in our case being the squared amplitude of the neurons, is applied. As the neurons represent the complex light field, this non-linearity is not only sufficient for optical neural networks, but outperforms linear classifiers  \cite{BPONN0,ONN3}. Physically, this is manifest in the form of optical detectors such as charge-coupled devices, which naturally measure the square amplitude of the complex field through photoelectric conversion. 

Any layers without a non-linearity represent a departure from traditional neural network architectures, as each layer's linear transformation can be viewed as a composition of all transformations, $\hat{T}_{\text{tot}}$, as $\hat{T}_{\text{tot}} = \hat{T}^0\hat{T}^1...\hat{T}^l$. However, unlike a traditional neural network, where the trainable parameters include all elements of a transformation matrix, $\hat{T}^l$, a DONN is limited in trainable parameters which form diagonal matrices as part of the construction of the full transformation matrix. Further, most networks are only focused on the phase changes, which places a further constraint in optimization and expressiveness of the architecture.  In addition, DONNs introduce additional degrees of freedom (DOF) such as inter-layer distances, spatial sampling rates, and nonlinearities, all of which influence the effective transformation. Due to the lack of a closed-form decomposition and the high dimensionality of trainable parameters, backpropagation is a particularly well-suited strategy allowing for a data-driven approximation of the optimal transformations. 

\subsection*{Backpropagation Algorithm}
At the core of machine learning are gradient descent and the backpropagation algorithm. The feed forward process involves encoding input data in the phase or amplitude channels of the first layer of neurons, $n^{l=0}_{i,j}$, and the prediction at the output of the optical neural network, $n^L_{i,j}$, can be compared to the labeled data set, $y^L_{i,j}$, using a cost function. The gradient descent algorithm determines the gradient of each trainable parameter and updates these parameters to reach a local minimum in the cost function. In multilayer networks, backpropagation computes the gradient at the final layer of neurons and propagates it backwards, adjusting each layer’s parameters based on their contribution to the gradient, thereby optimizing the network’s performance.

The trainable parameters for an optical architecture are the phase, $\phi^l$, for all layers in the neural network. For the derivation of the backpropagation algorithm, a two-layer neural network is assumed, meaning a composition of several linear transformations, $\hat{T^l}$, which encompass a total linear transformation $\hat{T}_{\text{tot}}$, and a single non-linearity dictated by the power square law. As multiple layers of linear transformations can be reduced to one layer in a typical neural network, for a DONN to be considered a deep neural network, additional non-linearities would be needed, which could be reconciled by encoding the output neurons detected by the camera into additional layers.  The feed-forward process is defined by:

\begin{subequations}\label{eq:ff}
\begin{align} 
\text{Linear Tranformation: }   \text{z}^l_\text{i,j} = \mathcal{F}^{-1}[\mathcal{F}[\text{n}^{l-1}_{k,l} \odot e^{i\phi_{k,l}^{l}}] \odot H_{k,l}^l] \\
   \text{Activation:   } \\
   \text{n}^l_\text{i,j} =  \text{z}^l_\text{i,j}  \text{ for } l \neq L \\
\text{n}^l_\text{i,j} =  |\text{z}^l_\text{i,j}|^2 \text{ for } l = L\\
\end{align}
\end{subequations}

where $l =1,\dotsc,L$. Notation follows the matrix form of the neurons rather than the vectorized form typically used in neural networks \cite{CompFO}. Here, $(i,j)$ and $(k,l)$ are layer-based indices which denote the column and row of all matrix elements, and $\odot$ refers to the Hadamard product, or element-wise multiplication. For simplicity, we consider every linear layer to have the same size. 

During training, the final layer can be compared to a target transformation by a cost function, which we define as mean squared error,
$\text{C}_\text{m} = \sum_{i,j=0}^{\text{N}} (\text{n}^{\text{L}}_{i,j} - y_{i,j})^2$. C$_{\text{m}}$ corresponds to the cost for a single image, where each image is described by index m. $\text{n}^{\text{L}}_{i,j}$ corresponds to the output prediction (imaged intensity), and $y_{i,j}$ corresponds to the target output. The goal is to determine the gradient of the cost function with respect to all weights, $\phi^l_{i,j}$ in all layers, $\nabla \text{C}_{\text{m}} (\phi^l_{i,j})$. We find that for any given layer the calculation of the gradient is:

\begin{subequations}\label{eq:bp_noe}
\begin{align} 
\text{For the last layer, } l = \text{L: }\\
 \frac{\partial \text{C}_{\text{m}}}{\partial z^{\text{L}}_{i,j}} = 4(n^{\text{L}}_{i,j}  - y_{i,j}) \odot (z^{\text{L}}_{i,j})\\
\frac{\partial \text{C}_{\text{m}}}{\partial \phi^L_{k,l}} =  \text{Re}[ i  n^{\text{L-1}}_{k,l} \odot e  ^{i\phi^L_{k,l}} \odot \mathcal{F}^{-1}[\mathcal{F}[\frac{\partial \text{C}_{\text{m}}}{\partial z^{\text{L}}_{i,j}} ] \odot  H_{k,l}^L ]] \\
\text{For all previous layers, } l  \text{: }\\
 \frac{\partial \text{C}_{\text{m}}}{\partial n^{l}_{k,l}} =  e  ^{i\phi^{l+1}_{k,l}} \odot \mathcal{F}^{-1}[\mathcal{F}[\frac{\partial \text{C}_{\text{m}}}{\partial z^{l+1}_{i,j}} ] \odot  H^{l+1}_{k,l} ] \\ \label{eq:partial}
      \frac{\partial \text{C}_{\text{m}}}{\partial \phi^{l}_{m,n}}  = \text{Re}[ i n^{\text{l-1}}_{m,n} \odot e  ^{i\phi^{l}_{m,n}} \odot \mathcal{F}^{-1}[\mathcal{F}[\frac{\partial \text{C}_{\text{m}}}{\partial n^{l}_{k,l}}] \odot  H_{m,n}^l ]]
\end{align}
\end{subequations}

For ease of understanding, we denote all matrices with the dummy variables $(i,j)$, $(k,l)$, and $(m,n)$, which correspond to the spatial indexing of the matrix elements. Thorough derivation of these equations is provided in the Supplementary Material. The equations do not require determination of the individual $\nabla \text{C}_{\text{m}}(\phi^l_{i',j'})$; rather, the gradient of all phase elements in a single layer can be computed at once. Note that when using the recursion for previous layers in Equation \ref{eq:partial},  $\frac{\partial \text{C}_{\text{m}}}{\partial z^{l+1}_{i,j}} = \frac{\partial \text{C}_{\text{m}}}{\partial n^{l+1}_{i,j}}$ due to the absence of a non-linearity. Additional non-linearities can be easily included by substituting the aforementioned gradient. 

Similarly to backpropagation algorithms in a typical electronic neural network, this can be further reduced if we denote an error between each layer \cite{DL_nielsen}. For a typical ANN, the error is given by $E^l = \frac{\partial \text{C}}{\partial z^l}$. However, due to the complex nature of an optical neural network, we include the additional phase term $e^{i\phi^l}$. If we denote error at any given layer as  $E^l_{k,l} =  e  ^{i\phi^l_{k,l}} \odot \mathcal{F}^{-1}[\mathcal{F}[\frac{\partial \text{C}_{\text{m}}}{\partial z^{\text{l}}_{i,j}} ] \odot  H_{k,l}^l ]] $, this can be reduced to:

\begin{subequations}\label{eq:bp_e}
\begin{align} 
\text{For the last layer, } l = \text{L: }\\
 \frac{\partial \text{C}_{\text{m}}}{\partial z^{\text{L}}_{i,j}} = 4(n^{\text{L}}_{i,j}  - y_{i,j}) \odot (z^{\text{L}}_{i,j})^*\\
 E^L_{k,l} =  e  ^{i\phi^L_{k,l}} \odot \mathcal{F}^{-1}[\mathcal{F}[\frac{\partial \text{C}_{\text{m}}}{\partial z^{\text{L}}_{i,j}} ] \odot  H_{k,l}^L ]] \\
\frac{\partial \text{C}_{\text{m}}}{\partial \phi^L_{k,l}} =  \text{Re}[ i n^{\text{L-1}}_{k,l} \odot E^L_{k,l}] \\
\text{For all previous layers, } l  \text{: }\\
 E^l_{m,n} =  e  ^{i\phi^{l}_{m,n}} \odot \mathcal{F}^{-1}[\mathcal{F}[E^{l+1}_{i,j}] \odot  H_{m,n}^l ] \\
      \frac{\partial \text{C}_{\text{m}}}{\partial \phi^{l}_{m,n}}  =  \text{Re}[ i n^{\text{l-1}}_{m,n} \odot E^l_{m,n} ]]
\end{align}
\end{subequations}

Interestingly, the results are nearly mathematically identical to the backpropagation algorithm for a typical neural network \cite{DL_nielsen}. The addition of the phase term to the error can be attributed to complex analysis. Following from commutativity between the derivative and the Fourier operators, which are linear, the main difference is that the errors and gradients themselves are decomposed into plane waves and ``propagated" utilizing the Fourier transform and it's inverse. 

\subsection*{Benchmarking on MNIST Database and Arbitrary Linear Transformations}
\begin{figure}
    \centering
    \includegraphics[width=1\linewidth]{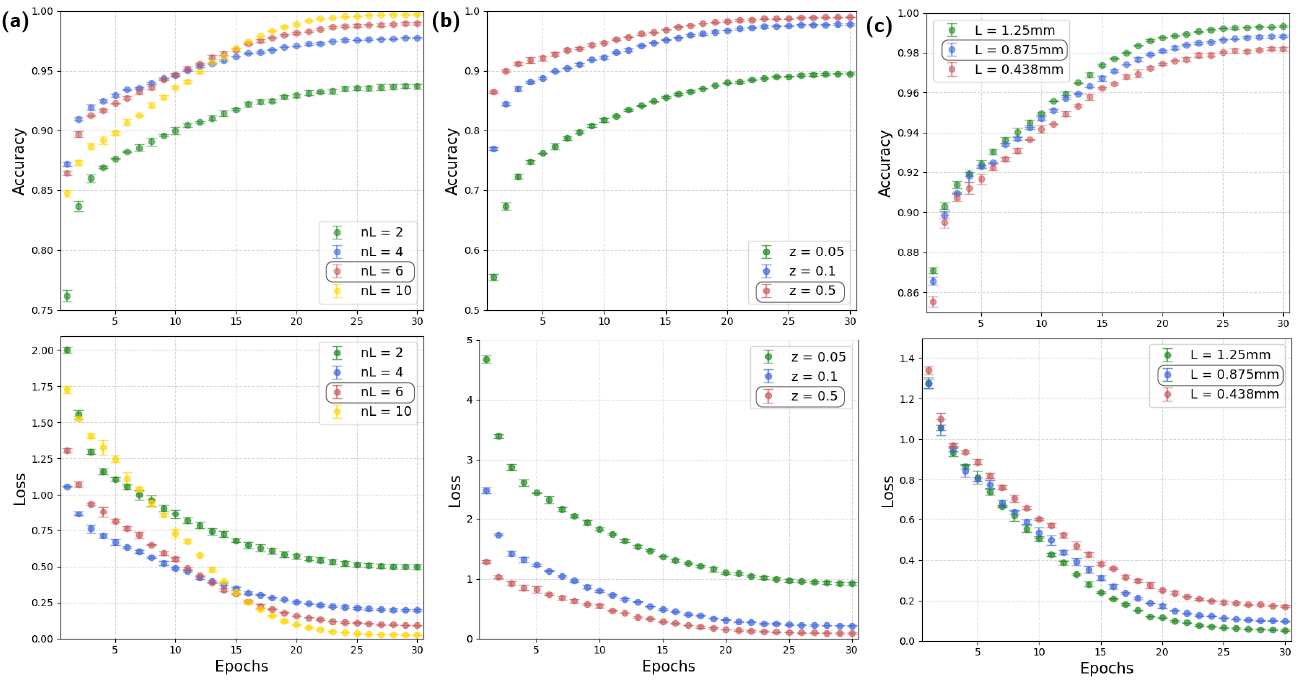}
    \caption{Training accuracy and loss over epochs for the classifying diffractive optical neural network architecture. The results were analyzed for the MNIST digits for varying (a) number of layers, (b) distance, z, in meters, and (c) detector region length. The model was trained with a cross categorical entropy loss for 30 epochs. The boxed legend entry in each figure indicates the fixed parameter value used for varying the other aformentioned parameters. The results are averaged over three trials, where the error bars constitute one standard deviation of the mean.}
    \label{fig:accloss}
\end{figure}

To test the computational efficiency and accuracy of our algorithm, we performed a classification task on the MNIST database of handwritten digits and several linear transformations \cite{NIST}. Beyond the trainable phase parameter, other DOFs such as the number of layers and inter-layer distances play a role in the classification accuracy for an optical architecture. The model was optimized with variable distance between layers, number of layers and detector region length for each class as shown in Figure \ref{fig:accloss}. 

Given the physical constraints of DONN architectures, several factors govern the learning capacity of our model. As shown in Figure \ref{fig:accloss} (a), additional phase-mask layers produce an increase in classification accuracy, enabling finer decision boundaries required for reliable classification. However, high numbers of circulant and diagonal factors, approaching theoretical upper bounds, are not required in practice to achieve high accuracy while remaining experimentally feasible. For an amplitude encoded optical neural network with six layers separated by 50 cm at each layer, we achieve a 98$\%$  and 97$\%$  training and testing set accuracy after 30 epochs. 

\begin{figure}[htbp!]
    \centering
    \includegraphics[width=1\linewidth]{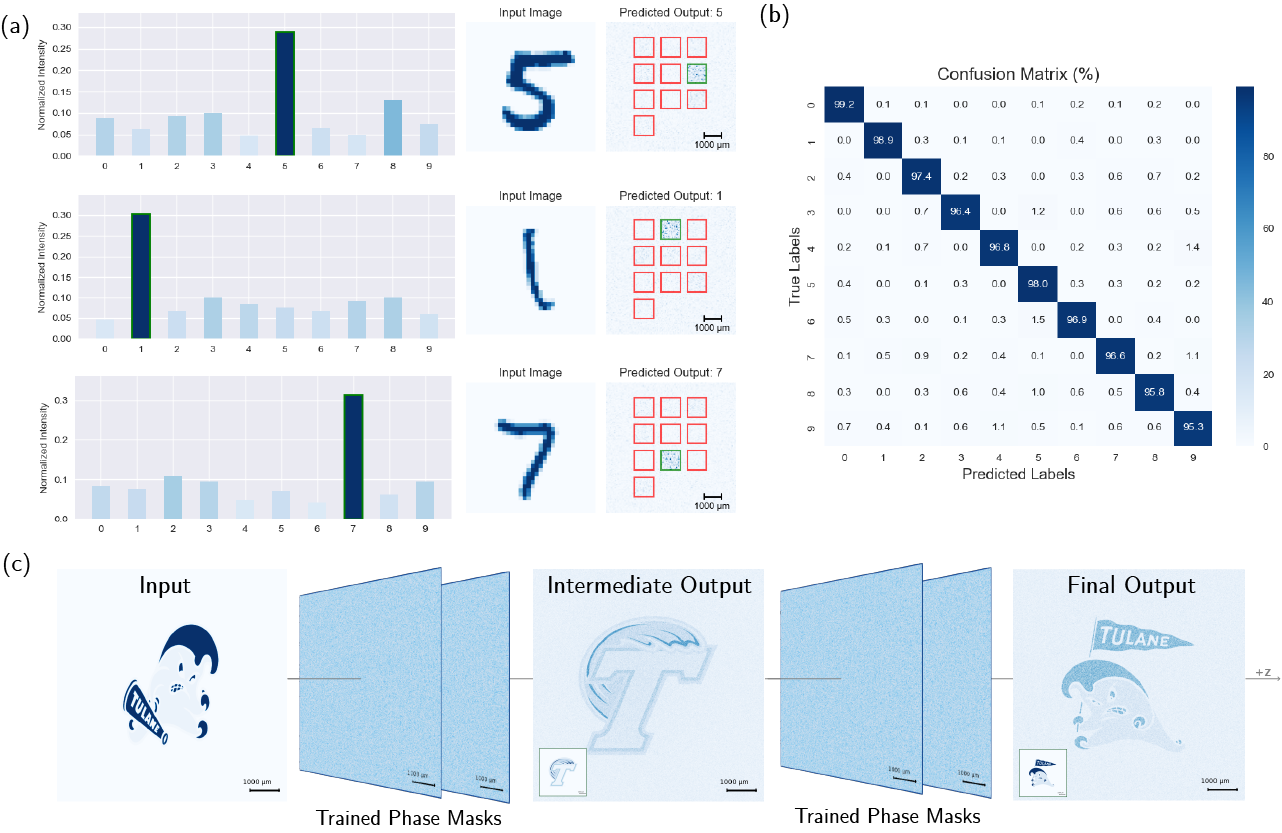}
    \caption{Testing results from the MNIST digits and the generation of arbitrary linear transformations. (a) Following the training, the classifying diffractive optical neural network (DONN) is used for the testing set for the MNIST digits. The normalized intensity at each detector region for the classifying DONN with six layers separated by 50cm is shown to the left. The detector region corresponding to the correct classification is outlined in green. The right displays the input intensity and the imaged final intensity for the respective MNIST digits. (b) Confusion Matrix for the test images given by percentage. (c) Generation of two linear transformations using our proposed algorithm given an initial phase encoded input and two target outputs. The intermediate output contains amplitude and phase information, and is used as input for the final output. Both transformations were achieved with negligible error utilizing only two phase masks. The goal output used to train the model is shown inset.}
    \label{fig:res}
\end{figure}

In addition to classification, the model can be utilized to generate arbitrary linear transformations. For this task, it learns phase masks that implement a target transformation directly in the spatial domain. We validate this capability using two encoding strategies -- initially the information is encoded in the phase of light, and an intermediate image where information is encoded in the amplitude and phase. The testing results for the MNIST handwritten digits and the generation of arbitrary linear transformations are shown in Figure \ref{fig:res}. 

For the generation of linear transformations between input and output field of views, only two trainable phase masks are needed. As these images are of size 1000 $\times$ 1000, this corresponds to 2 million adjustable parameters. Both target encodings converge using a mean squared error loss within 35 epochs, demonstrating the efficiency of our approach. This configuration offers an efficient way to generate arbitrary transformations given varying inputs, and is broadly applicable to use cases in wavefront shaping, image denoising, and image generation. 

\subsection*{Analysis of Computational Time}
\begin{figure}[htbp!]
    \centering
    \includegraphics[width=1\linewidth]{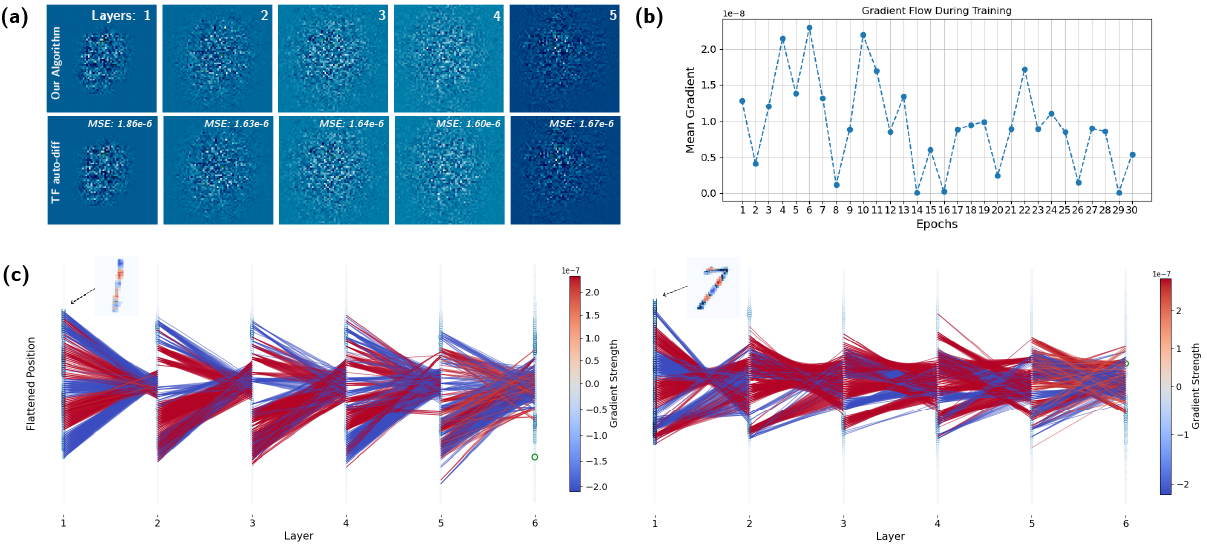}
    \caption{Analysis of the gradient for the MNIST handwritten digits. (a) Comparison of the gradients computed by our proposed algorithm and auto-differentiation done by Tensorflow (Google) after 5 epochs. The average mean-squared-error between the normalized gradients of both algorithms is shown inset. (b) The mean gradient over training for the 30 epochs of training using the proposed algorithm. We ensure evasion of vanishing gradient typically experienced during machine learning training. (c) The training images gradient between neurons for each layer. The intensity values in two-dimensions are flattened along the y-axis for each layer, and the corresponding two-dimensional representation and overlaid gradient heatmap is depicted inset for the first layer. For ease of view, only the top two-hundred normalized gradient connections for each layer are depicted using a heat map, with the gradient strength shown on the right.}
    \label{fig:gradients}
\end{figure}

Most relevant to our work are the gradient calculations and computational time. To ensure an appropriate assessment on training times, our results are compared to the widely utilized auto-differentiation technique to compute gradients. An identical classifying optical neural network was generated and trained using TensorFlow (Google), and the gradients were recorded. The resulting gradients after training both models with the same weight initialization, as well as gradient analysis, are shown in Figure \ref{fig:gradients} (a).

\begin{figure}[htbp!]
    \centering
    \includegraphics[width=1\linewidth]{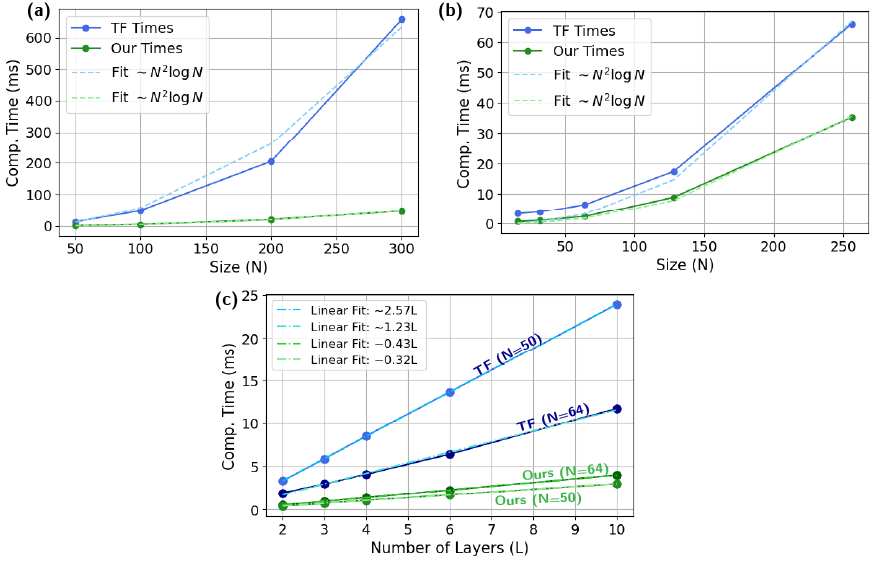}
    \caption{Computational time comparison between our implementation and auto-differentiation in TensorFlow. The computational time corresponds to the average time taken to process one image through forward and backward propagation.  Computational time in milliseconds for varying input image size, defined as ($\text{N}\times \text{N})$, for (a) by base 2 with powers from 4 to 8 and (b) non powers of two from $\text{N}$ = 50 to 300. Least squares scaling was used to compare the fit of the model  to the theoretical growth $\text{N}^2\text{log(N)}$ and establish a scale factor comparing the run time on both models. (c) Computational time in milliseconds for varying number of layers; a linear fit compared the growth in computational time of the model with respect to L. This was analyzed for $\text{N} = 2^6 = 64$ and $\text{N} = 50$. Tensorflow (Google) operations are most efficient with N that are powers of 2, while our algorithm does not have a preference.}
    \label{fig:comptime}
\end{figure}

The vanishing gradient problem, commonly observed in training of multi-layer machine learning models, is amplified when dealing with cascaded phase masks due to the necessity of multiple phase masks for a single linear transformation. With our algorithm, we ensure this problem is evaded using gradient stabilization techniques outlined in the Methods. We observe a stable decrease in the mean gradient over time, allowing for continued learning, which persists even for multi-layer architectures such as the ten layer DONN shown in Figure \ref{fig:accloss}. In addition to this gradient analysis, we observe nearly identical gradients with TensorFlow, with a mean squared error between both methods across all matrix elements to be on the order of $10^{-6}$. As both models follow an angular spectrum based feed-forward process, the theoretical computational complexity for the number of layers, L, and the size of the model, N, follows $\mathcal{O}(\text{L} \cdot\text{N}^2\log\text{N})$. The training time on average for an iteration of one image is shown in Figure \ref{fig:comptime}. Linear regression was used to approximate how the computation time scales with the number of layers. From the fitted slopes for both models, we found our proposed method to be approximately 8 times faster per image in the case of non power-two sized inputs and approximately 2.8 times faster for the case of base-two sized inputs. For varying size N, we fitted the dependence on the size of the model to $y = \text{N}^2\log\text{N}$ using least squares scaling. From this we were able to extract a scale factor that found the model was approximately 13.16 times faster for non-power of two sized inputs and 1.8 times faster for power of two size inputs, confirming consistent speed-up across both scaling dimensions. 

The approximate computational time per image was determined by dividing the total runtime by the number of images and training epochs. TensorFlow noticeably benefits from input sizes that are powers of two, which can be likely attributed to the fast Fourier transform, however its overall computational time remains higher than that of the proposed algorithm. In contrast, our method demonstrates consistent performance across all input sizes and exhibits no dependency on power-of-two dimensions. Our modeling was performed without any GPU acceleration, using Python version 3.11.5 on a 13 in. 2020 Macbook Pro laptop run on an Apple M1 chip with 8-core CPU (4 performance + 4 efficiency), 8 GB unified memory, macOS 14.6.1 (Sonoma). In comparison, current backpropagation algorithms proposed on a Nvidia TITAN XP GPU, Intel Xeon Gold 6126 CPU with 64 cores, 128GB RAM, Microsoft Windows 10 have taken approximately 3.8h and 5 hours for 15 epochs in Ref. \cite{ONN3} and Ref. \cite{BPONN1}, respectively. Reference  \cite{ONN0} utilized a GeForce GTX 1080 Ti GPU, Intel Core i7-7700 @ 3.60GHz, 64 GB RAM, Windows 10, using Python 3.5.0 and TensorFlow 1.4.0. and was reported to take 8 hours for 10 epochs. A more complex model, on similar computing hardware increased training time to 26 and 46 hours as discussed in Ref. \cite{ONNTimes}. 

Since the number of images, epochs, and image sizes can significantly affect \textit{in-silico} training times, we report our results in terms of per-iteration computational time. \textit{In-situ} training methods leverage the speed of light and optical parallelism to compute gradients scale-invariantly. This \textit{in-situ} approach, proposed by Ref. \cite{BPONN1}, achieved iteration times of approximately 80 milliseconds, limited primarily by the frame rates of current spatial light modulators and image sensors. Their reported \textit{in-silico} implementation required approximately 132 milliseconds per iteration for a 10-layer network for $\text{N}$=150. By contrast, our \textit{in-silico} six-layer optical neural network achieves a per-iteration time of approximately 10.5 milliseconds for the same size N, and achieves a faster computational time than the aforementioned \textit{in-situ} training for N=300. \textit{In-situ} approaches requires 4x upsampling for complex field generation modules, limiting the size of each layer to approximately N=250 without padding considerations. Given this, our training outperforms current gradient determination of \textit{in-situ }techniques in computational time. Further, as our metric to compute computational time corresponds to the total computational time divided by the number of images, and is computed without GPU acceleration, the computation times could be further decreased with more efficient computational power. 

By relaxing the requirement on an exact algebraic decomposition containing  $2\text{N}-1$ circulant and diagonal factors and instead fixing the number of factors to length L a-priori, we are able to achieve convergence for an experimentally realizable system. For the generation of an arbitrary linear transformation, the computational time to determine 2 million trainable parameters converged in approximately 10 seconds. For high resolution image generation, this fast computational time is paramount. In the context of machine learning, this relaxation is natural, as most classification tasks involve solving overdetermined systems, where no exact solution exists. To our knowledge, this is the first demonstration of a Fourier-based solution to this minimization problem, allowing for a low-factor cascaded phase mask model meeting the criteria of both experimental feasibility and computational efficiency. 

\section*{Discussion}
This work advances diffractive optical architectures as viable and scalable information processors by reducing the computational time associated with training. In traditional systems where a linear transformation is directly encoded in the optical set up, such as optical vector–matrix multiplication, the dimensionality of the linear transformation is constrained by the spatial resolution of current spatial light modulators or other diffractive elements. In contrast, cascaded phase mask-based architectures construct the transformation implicitly, and the phase-mask resolution governs instead the input vector dimensionality, enabling significantly larger-scale models within these existing hardware limits. Given the implicit nature of this transformation, however, designing the systems requires efficient methods for approximating arbitrary linear transformations using a limited number of circulant and diagonal factors. For this reason, data-driven methods are currently necessary to determine the required trainable parameters. As training machine learning models is inherently time-intensive, developing an efficient computational model is a key objective for high speed and low power statistical inference. 

Previous training algorithms for diffractive optical networks have either scaled poorly or failed to exploit the structured and sparse nature of the matrices underlying the linear transformations. In this work, we extend the computational capacity of generating such transformations by leveraging Fourier decomposition. By focusing solely on the trainable parameters, explicit construction of the overall $(\text{N}^2 \times \text{N}^2)$ transformation matrix associated with free-space propagation is bypassed, and the gradient of the trainable parameters is determined through element-wise multiplication in the Fourier domain. The algorithm was benchmarked on an MNIST digit classification task. We observed that high classification accuracy could be achieved using far fewer circulant and diagonal factors than the theoretical upper bound, maintaining both computational efficiency and experimental feasibility. This was further validated by successfully generating arbitrary linear transformations between various input and output images. Using the derivations for the gradients, gradient stability was improved during training through through the use of a logical detector layer, learning rate decay, and normalization to reduce loss and improve classification accuracy. Empirically, our method achieved an 8x speed up per layer in the case of non-power-of-two sized inputs. For larger-scale networks, necessary for high-resolution, the proposed backpropagation algorithm was approximately 13.16 times faster for non-power-of-two sized inputs. 

Due to the multitude of DOFs for task-specific diffractive optical neural networks, such as inter-layer distances, matrix size, number of layers, detector region size, etc., many simulations are typically needed before implementation on physical hardware. As DONN's increase in complexity to approach state-of-the-art classification performance, the use of the plane wave decomposition in both forward and backpropagation allows for necessary computational speed up. Further, efficient training aids in scaling DONNs and supports synthesis of large-scale linear optical transformations, or any signal processing which involves circulant and diagonal matrices. The scaling and high parallelism benefits that optical systems naturally provide, combined with implementation of faster training, in turn supports large-scale inference and other classical and quantum linear optical processing tasks. 

\section{Methods}
\subsection{Diffractive Optical Architecture Design and Training}
\subsubsection{Simulation using the angular spectrum method}
While there are many different ways to implement optical linear transformations, we focus on the composition of circulant and diagonal matrices, which can generate any arbitrary linear transformation \cite{OLTBase}. Mathematically, this transformation matrix, $\hat{T}^l$, can be decomposed into the following components,

\begin{subequations}\label{eq:3}
\begin{align} 
 \hat{T}^l = \hat{U}^l\hat{W}^l, \\
  \hat{W}^l = \text{diag}(e^{i\vec{\phi}^l}),\\
\hat{U}^l = \mathcal{\hat{F}}^{-1} \hat{\Lambda}^l \mathcal{\hat{F}} ,\label{eq:4} \\
\end{align}
\end{subequations}

which represents a phase transformation via \(\hat{W}^l\), followed by free-space propagation through \(\hat{U}^l\).  In the context of this work the transformation $\hat{W}^l$ assumes each diagonal element has a magnitude of one, which is suitable for optical devices such as spatial light modulators or other devices capable of directly modulating the wavefront's phase. The transformation \(\hat{U}^l\) describes free-space propagation using the angular spectrum method, where \(\mathcal{\hat{F}}\) and \(\mathcal{\hat{F}}^{-1}\) denote the discrete Fourier transform and inverse DFT operators, respectively. The term \(\hat{\Lambda}^l = \text{diag}(H)\) contains the transfer function,

\[
H = e^{ik\text{z}},
\]

where z is the propagation distance, $k_z = \sqrt{k^2-k_x^2-k_y^2}$,  $k=2\pi/\lambda$ and $\lambda$ is the wavelength. 

The Fourier transform can be interpreted as a decomposition of the input field, $n^0$,  into its plane wave components. Two-dimensional Fourier analysis allows for decomposition of a matrix in terms of their inner product with the basis vectors, orthogonal plane waves. The Fourier coefficients then represent the weight of each plane wave present in the original field. This representation is particularly powerful as plane waves are eigenfunctions of linear operators governing free-space propagation. The complex monochromatic light field, in our case characterized by the neurons, $\vec{n}^l_i$, is decomposed into orthogonal plane waves with a two-dimensional Fourier transform and propagated separately using solely element wise multiplication with the transfer function $H$. The resulting propagation at some distance away is the summation of the plane waves, which is done mathematically by the inverse Fourier transform. 

This change of basis allows for efficient computation. By way of illustration, consider the input field as a matrix of size $(\text{N}\times\text{N})$ undergoing a linear transformation. Typically, computing a linear transformation for an ANN involves vectorizing the two dimensional input utilizing matrix-vector multiplication scales as $\mathcal{O}(\text{N}^4)$. On the other hand, computing a Fourier Transform with the FFT algorithm scales $\mathcal{O}(\text{N}^2\text{log}(\text{N}))$ for a two dimensional input. This type of linear transformation, which is diagonalizable the the Fourier Transform, is a type of Toeplitz matrix structure called a circulant matrix, and is used widely in signal processing, sensing, solving ordinary and partial differential equations \cite{Circ0,Circ1,Circ2,Circ3}. Leveraging this change of basis offers a computationally efficient means of modeling light propagation and interactions through

\[
z^l_{i,j} = \mathcal{F}^{-1}[\mathcal{F}[n^{l-1}_{k,l} \odot e^{i\phi^l_{k,l}}] \odot H^l_{k,l}],
\]

where $(i,j)$ and $(k,l)$ corresponds to dummy variables indicating the element of the $(\text{N} \times \text{N})$ matrices describing the neurons, phase elements and transfer function, and $\odot$ corresponds to the Hadamard product. This approach calculates the transformation, \(\hat{T}^l\), implicitly, utilizing element-wise multiplication. 

\subsubsection{Pseudo-Code for the Diffractive Optical Architecture}
The pseudo-code for the feed forward process as described by Equation \ref{eq:ff} is shown in Algorithm \ref{alg:feedforward}.

\begin{algorithm}
\caption{Feedforward}
\label{alg:feedforward}
\begin{algorithmic}
\Require Complex Input neuron $n[0]$, number of layers $L$,  $H$, weights $\phi$
\Ensure Outputs $n$ for all layers

\For{$l \gets 1$ to $L+1$}
    \State $z[l] \gets \mathcal{F}^{-1}[\mathcal{F}[n[l-1] \odot e^{i\phi[l]}] \odot H]$ \Comment{Assuming phi-index starts at 1}
    \If{$l = L$}
        \State $n[l] \gets |z[l]|^2$
    \Else
        \State $n[l] \gets z[l]$
    \EndIf
\EndFor

\Return $n$ for all layers
\end{algorithmic}
\end{algorithm}

The process of determining $\nabla \text{C}_{\text{m}} (\phi^l)$ across all layers is repeated for all data in a given data set, or batch in the case of stochastic gradient descent. The gradient across all data is then averaged to $\nabla \text{C} (\phi^l) = \frac{1}{N} \sum^{N}_{m=1} \nabla \text{C}_{\text{m}} (\phi^l)$. The weights are updated as,

\begin{equation} \label{w_up}
\phi^l_{\text{new}} = \phi^l - \eta \nabla \text{C}(\phi^l),
\end{equation}

where $\eta$ corresponds to the learning rate. This gradient, $\nabla \text{C}(\phi^l)$, will determine the average change needed in the phase parameters in order to approach a minima. The completion of this process is called an epoch, and the process is repeated until a local minimum is reached. The pseudocode for determining the gradients is shown in \ref{alg:backpropagation}.

\begin{algorithm}
\caption{Gradient Calculation}
\label{alg:backpropagation}
\begin{algorithmic}
\Require Complex neurons $n$, Target $y$, number of layers $L$, $H$, weights $\phi$

\Comment{Last layer:}
\State $E[L] \gets e^{i\phi[L]} \odot \mathcal{F}^{-1}[\mathcal{F}[4(n[L] - y) \odot (z[L])^*] \odot H$
\State $\frac{\partial \text{C}}{\partial \phi}[L] \gets  \text{Re}[ i n[L-1] \odot E[L]]$

\Comment{All previous layers:}
\For{$l \gets L-1$ to $0$}
    \State $E[l] \gets e^{i\phi[l]} \odot \mathcal{F}^{-1}[\mathcal{F}[E[l+1]] \odot H]$
    \State $\frac{\partial \text{C}}{\partial \phi}[l] \gets  \text{Re}[ i n[l-1] \odot E[l]]$
\EndFor

\Return Gradients $\frac{\partial \text{C}}{\partial \phi}$ for all layers
\end{algorithmic}
\end{algorithm}

\subsubsection{Extending for Logical Detector Layer}
Our original analysis focuses on mean-squared-error as a loss metric. Due to the stringent constraints imposed by a mean-squared-error loss function for classification tasks, we generate a logical detector layer, which sums the intensity at each detector region, and a categorical cross-entropy loss was used for the classification task. This has been proven to stabilize optical neural network training, as well as increase final accuracy \cite{ONNloss}. Categorical cross-entropy can be defined as softmax activation and a cross-entropy loss for improved training accuracy and stability for multi-class classification \cite{pml}. In this approach, the output intensity from our DONN are converted into logits through a logical detector layer defined as

\[
\text{l}_k^{\text{L}} = \sum_{i,j} \text{n}^{\text{L}} \cdot \text{d}_k,
\]

where $k$ is the range of detector regions, $\text{d}_{k}$ defines the current detector region and acts as a logical mask, and the sampling space in the last layer $(i,j)$ is summed over. The logit values, $\text{l}^{\text{L}}_k$, were subtracted by the max value, $\text{max}(\text{l}^{\text{L}}_k)$, in order to prevent overflow, and the maximum value of the logits corresponds to the class prediction. The loss function is defined as

\begin{equation}
\text{C} = -\text{log} (\frac{e^{\text{l}^{\text{L}}_{y}}}{\sum_k e^{\text{l}^{\text{L}}_k}}).
\end{equation}

The gradient for all logits in a layer is given by

\begin{equation}
\frac{\partial \text{C}}{\partial \text{l}^{\text{L}}_k} = \frac{e^{\text{l}^{\text{L}}_{k}}}{\sum_k e^{\text{l}^{\text{L}}_k}} - y_k,
\end{equation}

where $y_k$ is the one-hot encoded true label. This gradient is then included in the chain rule derived for the full backpropagation gradient, defined in Equation \ref{eq:bp_noe}. 

\subsubsection{Further Gradient Stabilization}
Vanishing gradients is a common issue in any deep ANNs with multiple layers \cite{VanGrad}. As DONNs necessitate many layers just for a single overall linear transformation block, this difficulty becomes more apparent. Previous implementations have utilized types of rectifying linear units (ReLU) or sigmoid as an auxillary term, i.e. used in training but not in implementation \cite{ONNloss,TrainP0}. 

To train the classifying model, a learning rate decay was utilized with a decay rate of 0.99, and during backpropagation the gradients were normalized which we found improved the stability of the gradient over time. For the generation of arbitrary linear transformations, a higher learning rate was used with a decay rate of 0.98, and no normalization of the gradient was needed. Finally, an auxiliary function, $f$ was used to constrain the phase. As the phase is periodic, we found that a phase constraint given by the modulus,

\begin{equation}
    \phi^l_{i,j} = \phi^l_{i,j} \, \,  \text{mod} \, \, 2  \pi,
\end{equation}

was suitable, and was done after each batch. 

\subsection{Model Implementations}
\subsubsection{Aligning with Current Optical Devices}
To align with current optical devices, such as spatial light modulators, pixel pitch and resolution, we assumed the parameters for our model as shown in Table \ref{tab:optical_params}. 

\begin{table}[htbp!]
\centering
\begin{tabular}{lll}
\toprule
\textbf{Parameter} & \textbf{Classifying ONN}&\textbf{Arbitrary LTs}\\
\midrule
Distance ($z$) & [5, 10, 20, \textbf{50}, 100] cm  &70 cm\\
Wavelength ($\lambda$) & 795 nm  &795 nm\\
Field of View (FOV) length & 8 mm  &8 mm\\
Original image size & 28 $\times$ 28 pixels &Variable\\
Upsampling factor & 4  &None\\
Detector region length & [0.44, \textbf{.88}, 1.25] mm &--\\
Total array size ($N$) & 2$^7 \times $2$^7 = $128 $\times$ 128 &1000 $\times$ 1000\\
Effective pixel size ($\Delta x$) & 62.5 µm  &8 µm\\
Padding & 8 pixels per side  &200 pixels per side\\
\bottomrule
\end{tabular}
\caption{Optical simulation parameters used in the DONN and generation of arbitrary linear transformations. The bold values corresponds to the final parameter chosen.}
\label{tab:optical_params}
\end{table}

The field of view (FOV) length, which corresponds to the length in the $(x,y)$ plane that is sampled, was defined as 8 mm. The original 28 $\times$ 28 matrix given by the MNIST digits was upsampled by 4 to give an image of size 112 $\times$ 112 and padded by 8 pixels per side. Considering a typical pixel pitch for spatial light modulators is around 8$\mu$m, we assume an effective pixel size of 62.5 $\mu$m. As small detector regions have been shown to increase the overall classification accuracy, the detector region was reduced to 15 macropixels in length \cite{ONNDet}.  The total number of neurons is then N $\times$ N, where N = 128. For the arbitrary linear transformations, all images were resized to be 1000 $\times$ 1000 with padding of 200 pixels per side. 

\subsubsection{Training the Model}
The parameters for training the classifying model are shown in Table \ref{tab:nn_param}.
\begin{table}[htbp!]
\centering
\begin{tabular}{lll}
\toprule
\textbf{Parameter} & \textbf{Classifying ONN}&\textbf{Arbitrary LTs}\\
\toprule
Image size (NxN) & 128$\text{ x }$128  &1000 $\text{ x }$ 1000\\
\# Layers & [2,3,4,\textbf{6},10]  &3\\
\# Epochs & 30  &35\\
\# Parameters (x$10^3$)& [16, 33, 49, \textbf{82}, 148] &2000\\
Batch size & 10  &--\\
Learning rate & 0.05  &0.5\\
Optimizer& Adam  &Adam\\
Weight Constraint & mod 2$\pi$&mod 2$\pi$\\
Logical Det. Layer& Yes &No\\
Input Encoding& Amplitude  &Phase and Amplitude\\
Loss Metric & CCE &MSE\\
\bottomrule
\end{tabular}
\caption{Model parameters for \( z = 50\,\mathrm{cm} \). Bolded layer indicates the value used to report final figure results. Abbreviations: CCE - categorical-cross entropy, MSE - mean-squared error.}
\label{tab:nn_param}
\end{table}

The use of the logical detector region improved training dynamics compared to the previous MSE approach, increasing training and testing accuracy for an amplitude-encoded ONN to over 98$\%$ and 97$\%$, respectively. For the generation of arbitrary linear transformations, a MSE model was used. For this case, the model was tested using both amplitude/phase and phase only encoding. The initial layer assumed an incident Gaussian beam with radius 1.5 mm, and the input was the original image encoded in the phase, which can be achieved physically by a laser source incident on a spatial light modulator. The goal was then to achieve an intermediate output. The intermediate image had both phase and amplitude components, and this result was used as the input into the second model to generate a final output image. We found that both encoding methods were able to generate the goal output image with minimal loss. 

\bibliographystyle{ieeetr}
\bibliography{main}

\end{document}